\documentclass[10pt,journal,comsoc]{IEEEtran}
\usepackage{amsmath}
\usepackage{amsfonts}
\usepackage{amssymb}
\usepackage{bm}
\usepackage{algorithm}
\usepackage{float} 
\usepackage{algpseudocode}
\usepackage{xspace}
\usepackage{caption}
\usepackage{subcaption}
\usepackage{graphicx}
\usepackage{url}
\usepackage{caption}
\usepackage[noadjust]{cite}
\usepackage{graphicx}
\usepackage{multirow}
\usepackage{comment}
\usepackage{makecell}
\usepackage{ragged2e}
\usepackage{rotating}
\usepackage{lscape}
\usepackage[table]{xcolor} 
\usepackage{nomencl}
\usepackage{stfloats}
\makenomenclature
\usepackage{wrapfig}
\usepackage{makecell}
\usepackage{balance}
\usepackage[utf8]{inputenc}
\usepackage{ragged2e}
\usepackage{booktabs}

\usepackage{pifont}
\usepackage{tabulary}
\usepackage{comment}
\usepackage{lipsum}
\usepackage{xcolor}

\begin{document}
\title{Federated Learning for Edge Networks: Resource Optimization and Incentive Mechanism}
\author{Latif~U.~Khan$^\dagger$,~Shashi~Raj~Pandey$^\dagger$,~Nguyen~H.~Tran,\IEEEmembership{~Senior~Member,~IEEE,}~Walid~Saad,~\IEEEmembership{~Fellow,~IEEE},~Zhu~Han,~\IEEEmembership{~Fellow,~IEEE},~Minh~N.~H.~Nguyen,~and~Choong~Seon~Hong,~\IEEEmembership{~Senior~Member,~IEEE}

\IEEEcompsocitemizethanks{

\IEEEcompsocthanksitem L.~U.~Khan,~S.~R.~Pandey,~M.~N.~H.~Nguyen~and~C.~S.~Hong are with the Department of Computer Science \& Engineering, Kyung Hee University, Yongin-si 17104, South Korea.
\IEEEcompsocthanksitem N.~H.~Tran is with School of Computer Science, The University of Sydney, Sydney, NSW 2006, Australia.
\IEEEcompsocthanksitem Walid Saad is with the  Wireless@VT, Bradley Department of Electrical and Computer Engineering, Virginia Tech, Blacksburg, VA 24061 USA.
\IEEEcompsocthanksitem Zhu Han is with the Electrical and Computer Engineering Department, University of Houston, Houston, TX 77004 USA, and also with the Computer Science Department, University of Houston, Houston, TX 77004 USA, and the Department of Computer Science and Engineering, Kyung Hee University, South Korea.

}

\thanks{
	}}

\markboth{}{}%


\IEEEcompsoctitleabstractindextext{%
\justify
\begin{abstract} 
Recent years have witnessed a rapid proliferation of smart Internet of Things (IoT) devices. IoT devices with intelligence require the use of effective machine learning paradigms. Federated learning can be a promising solution for enabling IoT-based smart applications. In this paper, we present the primary design aspects for enabling federated learning at network edge. We model the incentive-based interaction between a global server and participating devices for federated learning via a Stackelberg game to motivate the participation of the devices in the federated learning process. We present several open research challenges with their possible solutions. Finally, we provide an outlook on future research.               
\end{abstract}

\let\thefootnote\relax\footnote{$^\dagger$ These authors contributed equally.}

}
\maketitle
\IEEEdisplaynotcompsoctitleabstractindextext
\IEEEpeerreviewmaketitle






\section{Introduction}
\setlength{\parindent}{0.7 cm}Emerging Internet of Things (IoT) applications such as augmented reality, autonomous driving, surveillance, and industry 4.0 generate significant amount of data. The effective deployment of such applications is thus reliant on the use of advanced machine learning techniques so as to properly exploit the generated data. However, traditional machine learning schemes use centralized training data at a data center which requires data transfer from a massive number of distributed IoT devices to a third-party location which raises serious privacy concerns and can be inefficient in its use of communication resources. To overcome these privacy and communication concerns, it is important to introduce distributed, edge-deployed learning algorithms such as \emph{federated learning} (\emph{FL}). FL allows privacy preservation by enabling distributed training without raw data transfer \cite{mcmahan2016communication}. \par

\begin{figure*}[!t]
	\centering
	\captionsetup{justification=centering}
	\includegraphics[width=13cm, height=13cm]{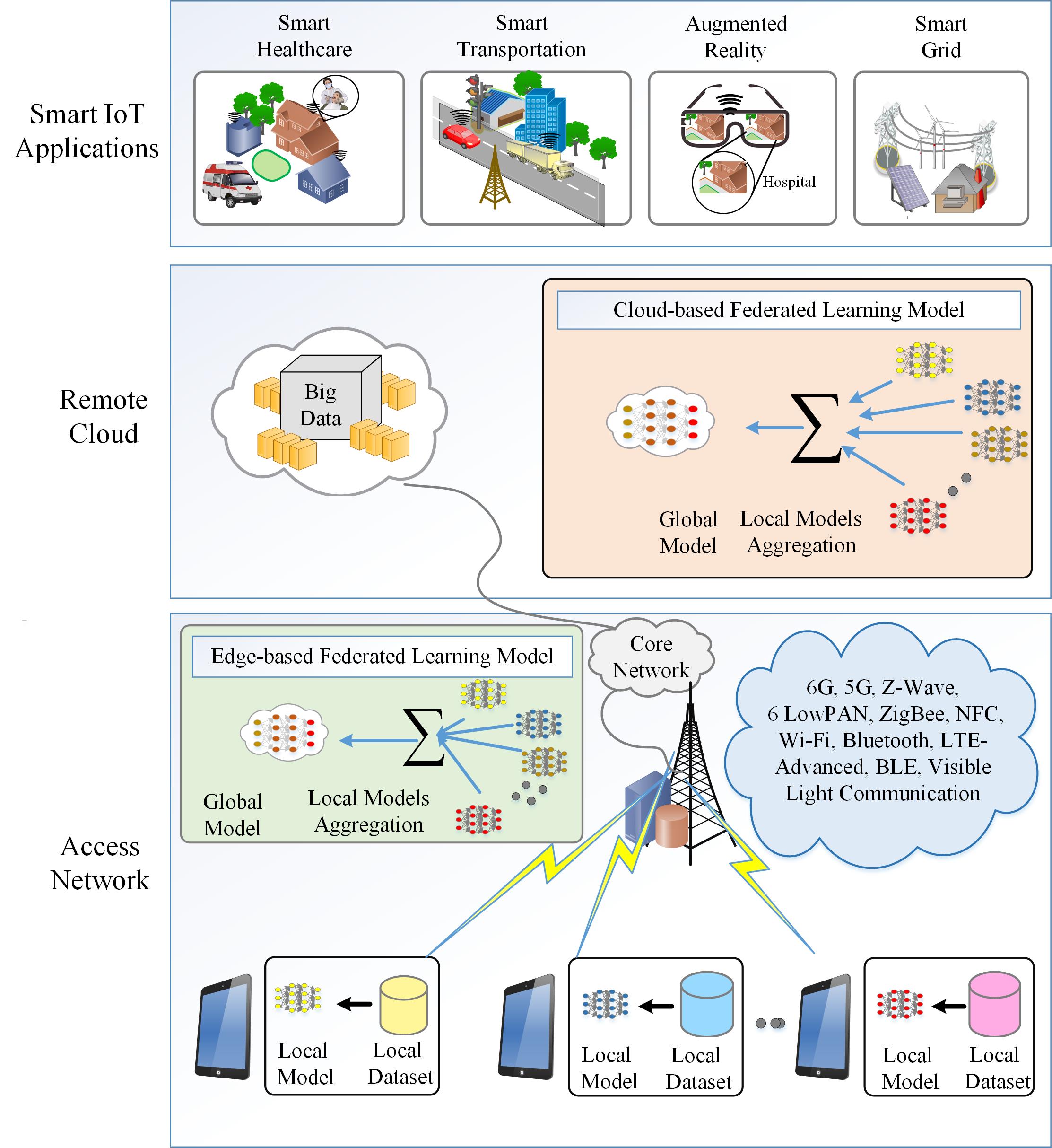}
	\caption{An overview of FL in enabling IoT-based smart applications}
	\label{fig:smartcityoverview}
\end{figure*}

\setlength{\parindent}{0.7 cm}An overview of how FL can enable IoT-based applications is presented in Fig.~\ref{fig:smartcityoverview}. To benefit from FL at the network edge, several challenges must be addressed that include resource management and incentive mechanism design to motivate the participation of users in the learning of a global FL model. Learning in the IoT has been studied in \cite{guo2018artificial,  chen2017caching,federated_survey, park2018wireless, Fed_edge1}. Works in \cite{guo2018artificial} and \cite{chen2017caching} rely on centralized learning solutions that have limited scalability and privacy-preservation. In \cite{federated_survey}, the authors presented the challenges of FL along with its existing solutions and applications in mobile edge network optimization. In \cite{park2018wireless}, the authors proposed an FL framework to provide efficient resource management at the network edge. However, the works in \cite{park2018wireless} and \cite{Fed_edge1} do not discuss the important challenges pertaining to incentive design and network optimization under edge-based FL. In contrast, the overarching goal of this article is to comprehensively review resource optimization and incentive mechanism for FL. In contrast to \cite {federated_survey} which focuses only at high-level challenges, we present a new perspective related to the development of incentive-based FL over edge networks using game theory. We also identify new challenges and open problems, different from \cite{federated_survey}. Our key contributions include:\par 
\begin{itemize}
    \item We present the key design aspects for implementing FL in edge networks.   
    \item  We present a Stackelberg game-based approach to develop an FL incentive mechanism. In this game, FL users can strategically set the number of local iterations to maximize their utility. Meanwhile, the base station (BS), acting as leader, uses the best response strategies of the users to maximize the FL performance. The BS's utility is modeled as a function of key performance metrics such as the number of global iterations and global accuracy level in the FL setting. 
    \item Finally, we present some key open research challenges along with guidelines pertaining to FL in edge networks. 
\end{itemize}\par
\section{Federated Learning at the Edge: Key Design Aspects}
\label{sec:designprinciples}
\subsection{Resource Optimization}
\setlength{\parindent}{0.7cm}Optimization of communication and computation resources is necessary to enable the main phases of FL local computation, communication, and global computation. When optimizing FL computational and communication resources, the original problem whose goal is to minimize the federated learning cost function can have a dual formulation without constraints. Moreover, if the original problem is convex, then dual problem has the same solution. Thus, the dual problem can be decoupled for obtaining a distributed solution in FL. Computation resources can be either those of a local device or of an edge server, whereas communication resources are mainly radio resources of the access network. In the local computation phase, every selected device iteratively performs a local model update using its dataset. The allocation of local device computational resources strongly depends on the device energy consumption, local learning time, and local learning accuracy. Further, the heterogeneity of the local dataset sizes significantly affects the allocation of local computational resources. Device energy consumption and local learning time are strongly dependent on the CPU capability. Increasing the device CPU frequency can increase the energy consumption and decrease the learning time. Similarly, the local computing latency increases for a fixed frequency with an increase in local learning accuracy. Evidently, there is a need to study the tradeoff between computation energy consumption, computational latency, learning time, and learning accuracy. Moreover, the access network and core network resources must be allocated optimally during the communication phase. 
\begin{figure*}[!t]
	\centering
	\captionsetup{justification=centering}
	\includegraphics[width=15cm, height=8cm]{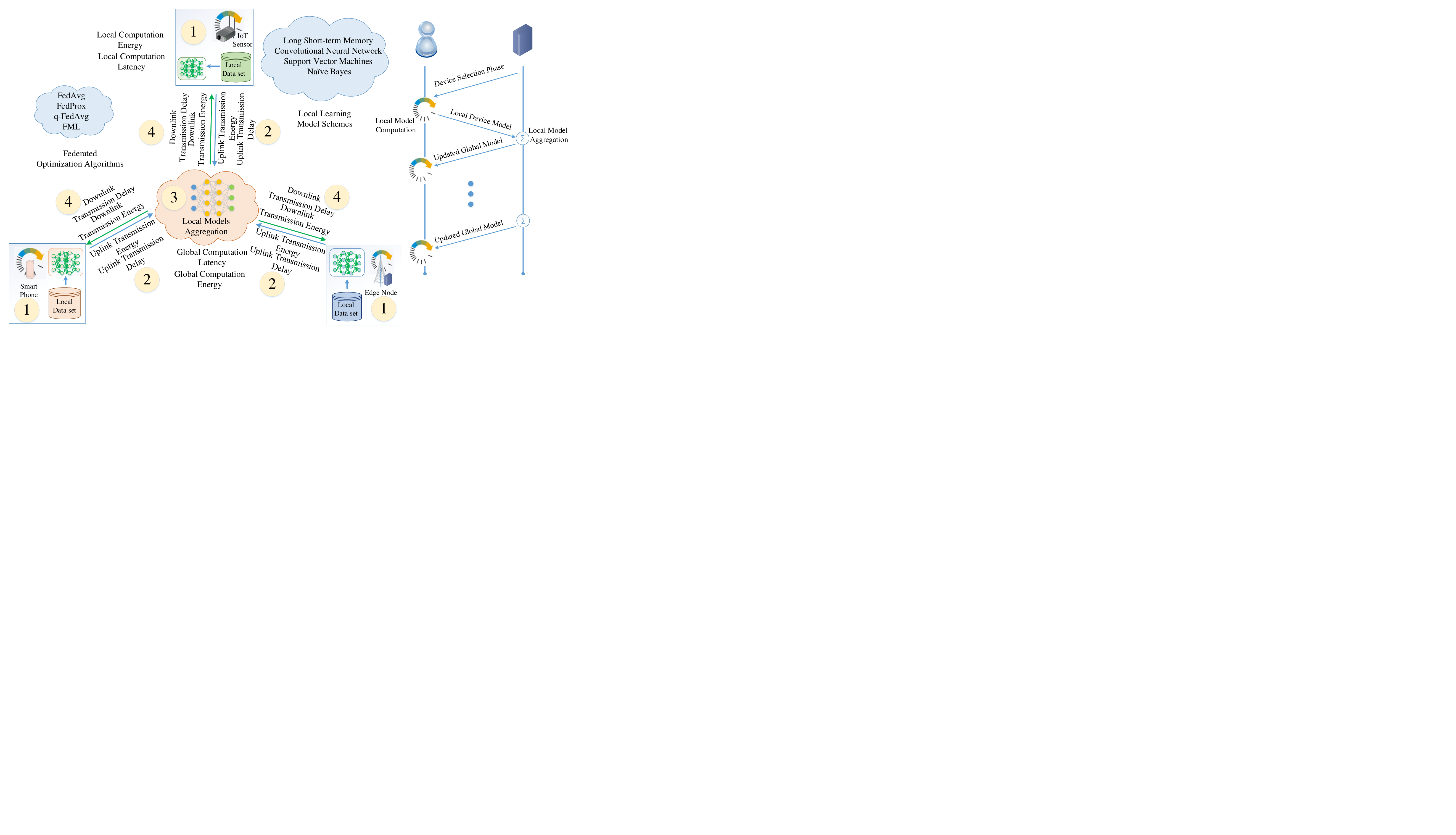}
	\caption{Federated learning sequence diagram}
	\label{fig:laerningsequencydiagram}
\end{figure*}
\subsection{Learning Algorithm Design}
\setlength{\parindent}{0.7cm}FL uses local and global computation resources along with communication resources. Several machine learning techniques, such as long short-term memory, convolutional neural network, and Naive Bayes schemes can be used at each local device. To enable FL, numerous optimization schemes, such as federated averaging (FedAvg) and FedProx can be used to train non-convex FL models \cite{fedAVG..}. FedProx is a modified version of FedAvg that captures both statistical and system heterogeneity among end-devices. FedAvg runs stochastic gradient descent (SGD) on a set of devices to yield local model weights. Subsequently, an averaging of the local weights is performed at the edge computing server located at BS. FedProx has similar steps as FedAvg, but the difference lies in local device minimizing of objective function that considers the objective function of FedAvg with an additional proximal term. By doing so, FedProx limits the impact of non-independent and identically distributed (non-IID) device data on the global learning model. FedAvg does not guarantee theoretical convergence, while FedProx shows theoretical convergence.\par
\setlength{\parindent}{0.7cm}In FedAvg and FedProx, all devices are weighted equally in global FL model computation without considering fairness, despite the differences in the device capabilities (e.g., hardware). To capture such fairness among devices, a so-called fairness enabled FedAvg algorithm was proposed \cite{li2019fair}. Fairness enabled FedAvg assigns higher weights to devices with poor performance by modifying the objective function of the typical FedAvg algorithm. To introduce potential fairness and reduce training accuracy variance, local devices having a high empirical loss (local loss function) are emphasized by assigning higher relative weight in the fairness enabled FedAvg. Meanwhile, in \cite{Fedlearning_edge_1} an adaptive control scheme was proposed to adapt the global FL aggregation frequency. This adaptive control scheme offers a desirable tradeoff between global model aggregation and local model update to minimize the loss function with resource budget constraint. All of the above-discussed methods are used for a single task global FL model. In real-world IoT systems, it is also of interest to use multi-task FL for handling multiple tasks, whose data is distributed among multiple edge nodes. A federated multi-task learning scheme was proposed in \cite{smith2017federated} by modifying the so-called communication-efficient distributed dual coordinate ascent (CoCoA) framework. To enable a wide variety of machine learning models, CoCoA supports objectives for linear reguarlized loss minimization \cite{jaggi2014communication}. In CoCoA, partial results from local computation are effectively combined using optimization problems primal-dual structure. In each round, CoCoA enables the use of any arbitrary optimization algorithm on a local dataset to solve a local learning problem by using distributed optimization for coping with system-level and statistical heterogeneity. \par  
\subsection{Hardware-Software Co-Design for Federated Learning}
\setlength{\parindent}{0.7cm}For a fixed hardware design, one can find optimal software design by searching for different architectures. However, this approach poses limitations on the design because neural network design is strongly dependent on the used dataset. Therefore, there is a need to jointly consider both hardware design space and neural architecture search space for a more flexible design of the end-device for FL \cite{software_hardware_design}. One promising approach for efficient design of end-devices involved in FL is hardware-software co-design. Several approaches such as high-level synthesis, co-verification-based embedded systems, and virtual prototyping can be used for hardware-software co-design of IoT devices. A design based on the virtual prototyping uses computer-aided engineering, computer-automated design, and computer-aided design for the validation of a design before prototype implementation, whereas a high-level synthesis offers an automated design process by creating digital hardware based on the algorithmic description for the desired behavior. The prominent challenges of high-level synthesis-based design are wired signal and multiplexer delays. Moreover, co-verification-based embedded systems enable concurrent testing and debugging of both software and hardware design, however, such designs require successful interactions between hardware and software teams.
\subsection{Incentive Mechanism Design}
\setlength{\parindent}{0.7cm}The design of mechanisms that incentivize users to participate in FL is a key challenge. Incentives are possible in different forms, such as user-defined utility and money-based rewards. Several frameworks such as game theory and auction theory can be used in the design of FL incentives \cite{han2012game, rajpandey1}. One can design an incentive mechanism using game theory while considering both communication and computation costs. The communication cost can be defined as the total number of rounds used for the interactions between the edge server and end-devices, whereas the computational cost can be the number of local iterations required to compute the local learning model \cite{federated_survey}. For synchronous aggregation, given a fixed number of global FL rounds between end-devices and edge server, the convergence rate of the global FL model has a proportional relationship with the number of local iterations. An increase in the number of local iterations minimizes the local learning model error and thus, few global FL rounds are required to reach a certain global FL model accuracy. Therefore, for a fixed global FL model accuracy, an increase in computational cost reduces communication cost and vice versa. For instance, consider a incentive mechanism game whose players are the edge server and edge users. The edge server announces a reward as an incentive to the participating users while maximizing its benefits in terms of improving global FL model accuracy. Meanwhile, the edge users maximize their individual utilities to improve their benefit. One example of a user utility could be the improvement of local learning model accuracy within the allowed communication time during FL training. An improvement in the local learning model accuracy of the end-user increases its incentive from the edge server and vice versa. This process of incentive-based sharing of model parameters continues until convergence to some global model accuracy level. \par      
\section{Incentive Based Federated Learning Over Edge Networks}
\subsection {System Model}
\setlength{\parindent}{0.7cm}Consider a multi-user system comprised of a BS and a set of user devices with non-IID and heterogeneous data sizes. Enabling FL over such edge networks involves the use of the computational resources at both device and cloud levels, as well as network communication resources. In a typical FL environment, participating user equipment (UE) must iterate over their local (possibly non-IID) data to train a global model. However, UEs are generally reluctant to participate in FL due to limited computing and communication resources. Thus, enabling FL requires some careful design considerations:
\begin{itemize}
\item First, to motivate UEs for participation, it is necessary to model the economic interaction between the BS  and the UEs. Within each global iteration, the BS can offer a reward rate (e.g., \$/iterations) to the UEs for selecting the optimal local iteration strategy (i.e., CPU-frequency cycle) that can minimize the overall energy consumption of FL, with a minimal learning time. 
\item The set of resource-constrained UEs involved in FL has numerous heterogeneous parameters: Computational capacity, training data size, and channel conditions. This heterogeneity significantly affects the local learning model computation time for a certain fixed local model accuracy level. For a synchronous FL setting, the local learning model accuracy will be different for different UEs due to both data and system heterogeneity. Therefore, it is necessary to tackle the challenge of heterogeneous local learning model accuracy for the UEs in synchronous FL. 
\item One approach for handling the communication-computation tradeoff in FL is via an appropriate client selection strategy. Selecting the IoT devices with sufficient computing power and training data, jointly improves FL model accuracy and training costs \cite{federated_survey}. In our previous work \cite{dinh2019federated}, we jointly optimized the computing time and energy consumption of FL over wireless networks. The problem studied in \cite{dinh2019federated} captures two tradeoffs: (a) UE energy consumption and FL time via variations in device CPU-cycles/sec and (b) computational and communication latencies for FL accuracy. However, here, we use a Stackelberg game-based incentive mechanism to select a set of IoT devices willing to join the model training process. Then, the selected set will collaboratively train a global model while minimizing the overall training costs, i.e., computation and communication cost.
\end{itemize}
\subsection{Stackelberg Game Solution}
\setlength{\parindent}{0.7cm}The BS employs an incentive mechanism for motivating the set of UEs to participate in global FL model training. However, heterogeneous UEs have different computational and communication costs for training and, thus, they expect different rewards. Moreover, the BS seeks to minimize the learning time while maximizing the accuracy level of the learning model. This complex interaction between the BS and the UEs can be cast as a Stackelberg game with one leader (BS) and multiple followers (UEs). For the offered reward, the BS maximizes its utility modeled as a function of key FL performance metrics such as the number of communication rounds needed to reach a desirable global FL model accuracy level. Correspondingly, the UEs will respond to the BS-offered reward and choose their local iteration strategy (i.e., select a CPU-frequency cycle for local computation) to maximize their own benefits \cite{rajpandey1}. Evaluating the responses from the UEs, the BS will adjust its reward rate, and the process repeats until a desired accuracy level is obtained. To this end, the BS must design an incentive mechanism to influence available UEs for training the global model. In this framework, the sequence of interactions between the BS and the UEs to reach a Stackelberg equilibrium is as follows:
\begin{itemize}
    \item Initially, each UE submits its best response (i.e., optimal CPU-frequency) to the BS for the offered reward rate, to maximize its local utility. Specifically, each UE considers the viability of the offered reward rate for their incurred computational and communication costs in FL.
    \item Next, the BS evaluates these responses, updates the global model, and broadcasts its offered reward rate to the UEs to maximize its own utility function. The utility of the BS is modeled as strictly concave function of key FL performance metrics such as the number of global iterations required to reach global accuracy for a given local relative accuracy. 
    

    \item Given the optimal offered reward, the UEs will correspondingly tune their strategy and update response that solves their individual utility maximization problem. This iterative process continues in each round of interaction between the BS and UEs.
    
\end{itemize} 
\setlength{\parindent}{0.7cm}In summary, we follow the best response algorithm to achieve the Stackelberg equilibrium. For this, with the first-order condition, we first find a unique Nash equilibrium at the lower-level problem (among UEs), and, then, use a backward induction method to solve the upper-level BS problem.\par

\begin{figure*}[t!]
	\centering
	\begin{subfigure}[b]{0.3\textwidth}
		\includegraphics[width=2.2in]{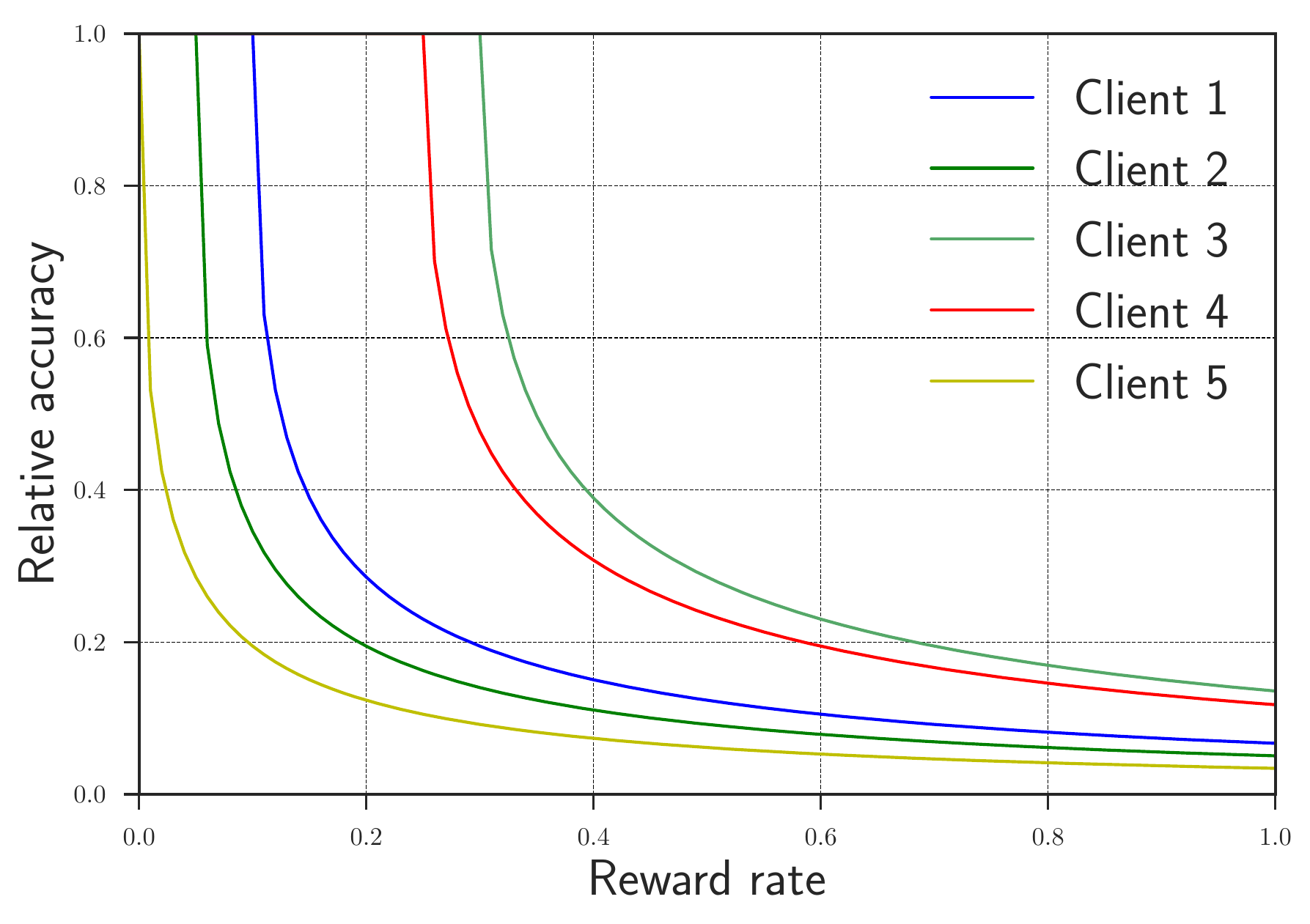}
		\caption{}
	\end{subfigure}%
    	\hfill
	\begin{subfigure}[b]{0.3\textwidth}
		\includegraphics[width=2.2in]{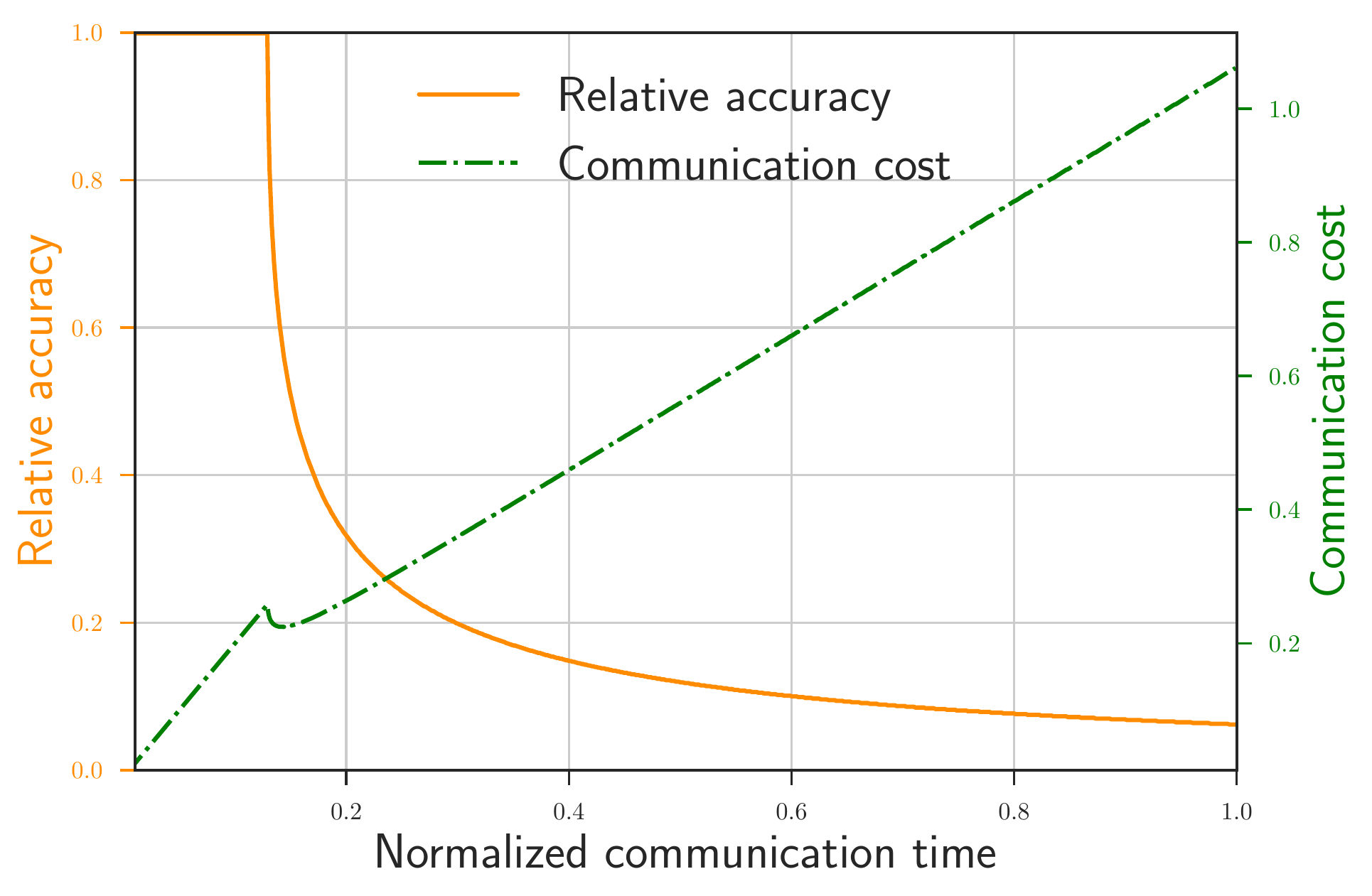}
		\caption{}	
	\end{subfigure}
		\hfill
	\begin{subfigure}[b]{0.3\textwidth}
		\includegraphics[width=2.2in]{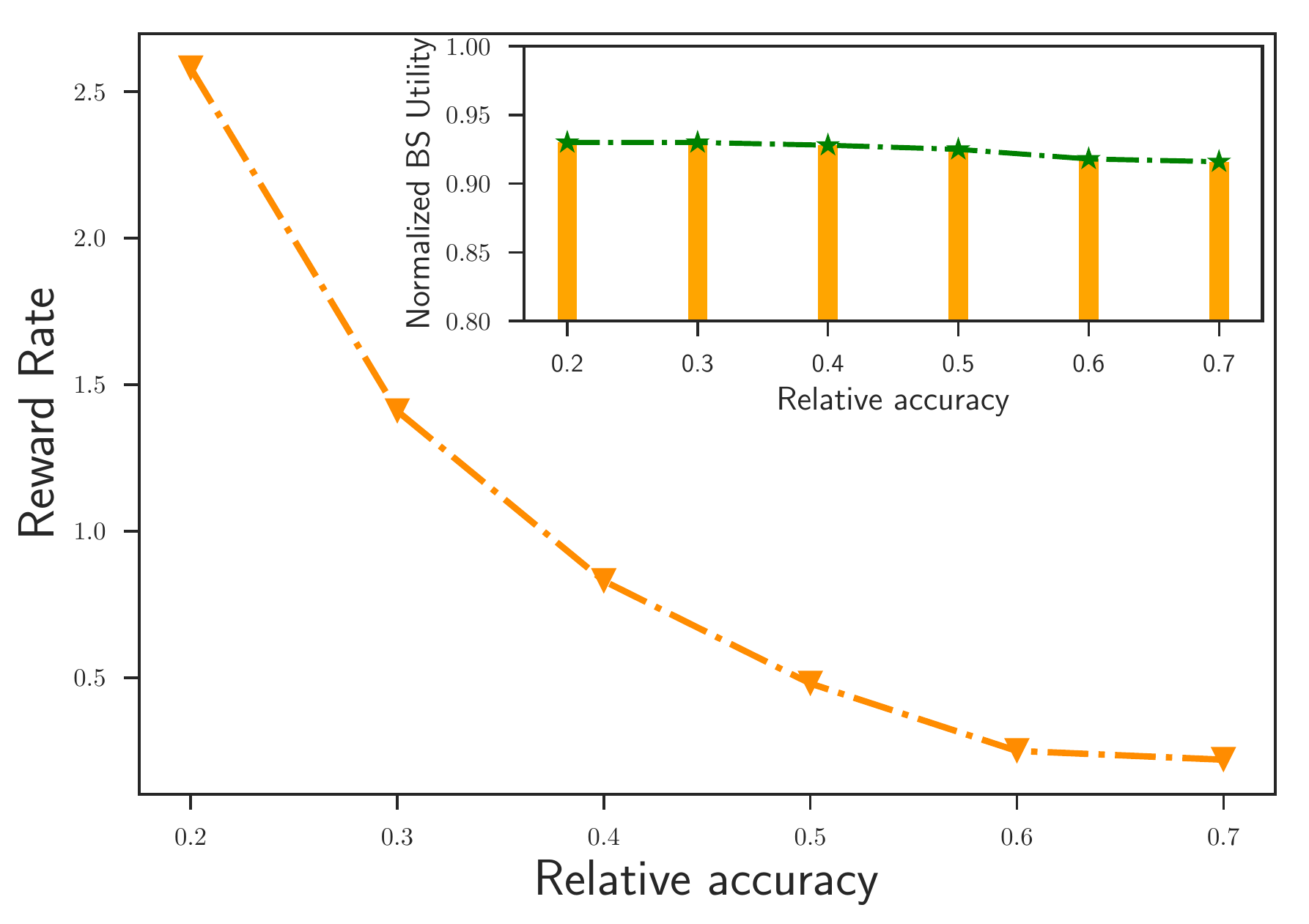}
		\caption{}	
	\end{subfigure}
	\centering
	\caption{Impact of (a) offered reward rate on client's (UEs) iteration strategy for corresponding relative local accuracy, (b) communication time with relative accuracy, (c) offered reward rate, normalized BS utility versus local relative accuracy.}
	\label{fig:response}
\end{figure*}
\subsection{Performance Evaluation}
\setlength{\parindent}{0.7cm}We now evaluate the performance of our incentive-based FL model by examining the contributions of each FL-participating UE. We investigate the impact of communication channel conditions and local computational characteristics on the accuracy of global federated learning model. We evaluate the impact of the offered reward in terms of communication cost versus local relative accuracy to characterize the system performance in FL. For FL, we adopt a classification task using multinomial logistic regression and distribute the MNIST dataset among participating UEs \cite{mcmahan2016communication}. For federated optimization, we use the modified CoCoA framework \cite{smith2017federated}. The distributed federated optimization scheme of \cite{smith2017federated} allows us to tackle both system-level and statistical heterogeneity efficiently. We consider five participating UEs having different channel conditions and an equal local data size. At each UE, we define the mean square error of the learning problem, i.e., the local relative accuracy metric. For the UEs utility, we choose a concave function of the local relative accuracy and the BS-offered reward.\par
\setlength{\parindent}{0.7cm}In  Fig.~\ref{fig:response}a, the impact of the offered reward rate on the relative accuracy for five UEs is shown. The accuracy improves when the relative accuracy value (x-axis) is smaller. Intuitively, an increase in the offered reward rate will motivate UEs to iterate more within one global iteration, resulting in a better accuracy. The heterogeneous UE responses is the result of individual computational limitations, local data size, and communication channel conditions. The impact of the communication channel conditions on local relative accuracy for a randomly chosen UE, with defined computational characteristics and local data size is illustrated in Fig.~\ref{fig:response}b. For clarity, we use a normalized communication time to quantify the adversity of channel conditions. Here, a unit value for the normalized communication time signifies poor channel conditions. As the communication time increases, the UEs perform more local iterations to avoid expensive communication costs. Fig.~\ref{fig:response}c presents the relationship between the offered reward rate and the local relative accuracy at the UEs. The offered reward rate reveals the optimal response of the UEs that maximizes their own utilities for given channel conditions. Here, we have consistency in the normalized BS utility function for various response behaviors of the UEs to the offered reward rate. Thus, it is crucial to have an appropriate incentive design to align the responses of the participating UEs for improving the FL performance.\newline
\section{Open Research Challenges}
\subsection{Resource Optimization for Blockchain based Federated Learning} 
\setlength{\parindent}{0.7cm}An attacker might attack the centralized FL server in order to alter global model parameters. In addition, a malicious user might alter FL parameters during communication. To cope with such security and robustness issues, blockchain based FL (BFL) can be used. BFL does not require central coordination in the learning of the global model which yields enhanced robust operation. In BFL, all users send their local model parameters to their associated miners, which are responsible for sharing local model updates through a distributed ledger. Finally, local model updates of all the devices involved in learning are sent back by miners to their associated devices for the local models aggregation. Although BFL provides benefits of security and robustness, it faces significant challenge of computational and communication resource optimization to reach a consensus among all miners. Static miners can be implemented at the BS, whereas wireless mobile miners can be implemented using drones. However, drone-based mobile miners pose more serious resource allocation challenges than static miners at the BS.    
\subsection{Context-Aware Federated Learning}
\setlength{\parindent}{0.7cm}
How does one enable more specialized FL according to users contextual information? Context-awareness is the ability of a devices/system to sense, understand, and adopt its surrounding environment. To enable intelligent context-aware applications, FL is a viable solution. For instance, consider keyboard search suggestion in smartphones in which the use of FL is a promising solution. In such type of design, we must consider context-awareness for enhanced performance. Unique globally shared FL model must be used separately for regions with different languages to enable more effective operation. Therefore, the location of the global model must be considered near that region (i.e., micro data center) rather than a central cloud. \par
\subsection{Mobility-Aware Federated Learning} 
\setlength{\parindent}{0.7cm}How does one enable seamless communication of smart mobile devices with an edge server during the learning phase of a global FL model? A seamless connectivity of the devices with a centralized server during the training phase must be maintained. Mobility of devices must be considered during the device selection phase of FL protocol. Deep learning-based mobility prediction schemes can be used to ensure the connectivity of devices during FL training.\par
\section{Conclusions and Future Recommendations}
\setlength{\parindent}{0.7cm}In this paper, we have presented the key design aspects, incentive mechanism, and open research challenges, for enabling FL in edge networks. We have identified four key design aspects: resource optimization, incentive mechanism, learning algorithm design, and hardware-software co-design-based end-devices, for FL at the network edge. We have shown that game-theoretic incentive mechanisms can be used to effectively model interaction between devices and edge server for FL. This work can potentially make FL amenable for implementation in diverse 5G-enabled smart IoT applications such as intelligent transportation systems, industry 4.0, and digital health care. Finally, we present several recommendations for future research: 
\begin{itemize}
\item Generally, FL involves training of a global FL model via an exchange of learning model updates between a centralized server and geographically distributed devices. However, wireless devices will have heterogeneous energy and processing power (CPU-cycles/sec) capabilities. Some of the devices might have noisy local datasets. Therefore, there is a need for novel FL protocols that will provide criteria for the selection of a set of local devices having sufficient resources. The selection criteria of the devices must include long-lasting backup power, sufficient memory, accurate data, and higher processing power. 
\item A set of densely populated devices involved in FL might not be able to have real-time access to the edge server located at the BS due to a lack of communication resources. To cope with this challenge, one can develop new FL protocols based on socially-aware device-to-device (D2D) communication. Socially-aware D2D communication has an advantage of reusing the occupied bandwidth by other users while protecting them by keeping the interference level below the maximum allowed limit. Initially, multiple clusters based on social relationships and the distance between devices should be created. Then, a cluster head is selected for every cluster based on its highest social relationship with other devices. Within every cluster, a sub-global FL model is trained iteratively by exchanging the model parameters between the cluster head and its associated devices. Then, the sub-global FL model parameters from all cluster heads are sent to the BS for global model aggregation. Finally, the global FL parameters are sent back to cluster heads which in turn disseminate them to their associated cluster devices.        

\item Exchange of learning model updates via blockchain offers enhanced security. However, reaching consensus via traditional consensus algorithms among blockchain nodes can add more latency to the learning time. Therefore, it is recommended to design novel consensus algorithms with low latency.



\end{itemize}



\bibliographystyle{IEEEtran}
\bibliography{Database}

\begin{IEEEbiography}{Latif U. Khan} is pursuing his Ph.D. degree in Computer Engineering at Kyung Hee University (KHU), South Korea. He received his MS (Electrical Engineering) degree with distinction from University of Engineering and Technology (UET), Peshawar, Pakistan in 2017. His research interests include analytical techniques of optimization and game theory to edge computing, federated learning, and end-to-end network slicing.  
\end{IEEEbiography}

\begin{IEEEbiography}{Shashi Raj Pandey } is currently pursuing his Ph.D. degree  from Computer Engineering Department at Kyung Hee University (KHU), South Korea. He received the B.E degree in Electrical and Electronics with specialization in Communication from Kathmandu University, Nepal in 2013. His research interests include network economics, game theory, wireless communications and networking, edge computing, and machine learning.
\end{IEEEbiography}

\begin{IEEEbiography}{Nguyen H. Tran}(S'10-M'11-SM'18) is currently working as a senior lecturer in School of Computer Science, The University of Sydney. He received the BS degree from Hochiminh City University of Technology and Ph.D. degree from Kyung Hee University, in electrical and computer engineering, in 2005 and 2011, respectively. His research interest is to applying analytic techniques of optimization and game theory to cutting-edge applications.
\end{IEEEbiography}

\begin{IEEEbiography}{Walid Saad } (S’07, M’10, SM’15, F’19) received his Ph.D. degree from the University of Oslo in 2010. Currently, he is a professor in the Department of Electrical and Computer Engineering at Virginia Tech. His research interests include wireless networks, machine learning, game theory, cybersecurity, unmanned aerial vehicles, and cyber-physical systems. He is the author/co-author of eight conference best paper awards and of the 2015 IEEE ComSoc Fred W. Ellersick Prize.
\end{IEEEbiography}

 \begin{IEEEbiography}{Zhu Han} (S’01, M’04, SM’09, F’14) received his Ph.D. degree in electrical and computer engineering from the University of Maryland, College Park. Currently, he is a professor in the Electrical and Computer Engineering Department as well as in the Computer Science Department at the University of Houston, Texas. Dr. Han is an AAAS fellow since 2019. Dr. Han is 1\% highly cited researcher since 2017 according to Web of Science.

\end{IEEEbiography}

\begin{IEEEbiography}{Minh N. H. Nguyen } is currently pursuing his Ph.D. degree from Computer Engineering at Kyung Hee University (KHU), South Korea. He received the BE degree in Computer Science and Engineering from the Hochiminh City University of Technology in 2013. His research interest includes Network Optimization, Mobile Edge Computing, Internet of Things.

\end{IEEEbiography}

\begin{IEEEbiography}{Choong Seon Hong} (S’95-M’97-SM’11) is working as a professor with the Department of Computer Science and Engineering, Kyung Hee University. He is currently an Associate Editor of the Future Internet, International Journal of Network Management, and Journal of Communications and Networks and an Associate Technical Editor of the IEEE Communications Magazine. His research interests include machine learning, edge computing, future internet and UAV networks.
\end{IEEEbiography}


\balance
\end{document}